\documentclass{article}

\usepackage{arxiv}

\usepackage[utf8]{inputenc} 
\usepackage[T1]{fontenc}    
\usepackage{hyperref}       
\usepackage{url}            
\usepackage{booktabs}       
\usepackage{amsfonts}       
\usepackage{nicefrac}       
\usepackage{microtype}      
\usepackage{lipsum}		
\usepackage{graphicx}
\usepackage{natbib}
\usepackage{doi}
\usepackage{makecell} 

\title{Performance of OpenBCI EEG Binary Intent Classification with Laryngeal Imagery}

\author{Samuel Kuhn \\
	Department of Psychology and Neuroscience\\
	Regis University\\
	Denver, CO 80221 \\
	\texttt{skuhn002@regis.edu} \\
	\And
	Nathan George \\
	Data Sciences Department\\
	Regis University\\
	Denver, CO 80221\\
	\texttt{ngeorge@regis.edu} \\
}

\renewcommand{\shorttitle}{\textit{arXiv} Template}

\hypersetup{
pdftitle={Performance of OpenBCI with Laryngeal Imagery},
pdfauthor={Samuel Kuhn, Nathan George},
}

\begin{document}
\maketitle

\begin{abstract}
One of the greatest goals of neuroscience in recent decades has been to rehabilitate individuals who no longer have a functional relationship between their mind and their body. Although neuroscience has produced technologies which allow the brains of paralyzed patients to accomplish tasks such as spell words or control a motorized wheelchair, these technologies utilize parts of the brain which may not be optimal for simultaneous use. For example, if you needed to look at flashing lights to spell words for communication, it would be difficult to simultaneously look at where you are moving. To improve upon this issue, this study developed and tested the foundation for a speech prosthesis paradigm which would utilize the innate neurophysiology of the human brain's speech system. In this experiment, two participants were asked to respond to a yes or no question via an EEG-based BCI of three different types; SSVEP-based, motor imagery-based, and laryngeal-imagery-based. By comparing the accuracy of the two established BCI paradigms to the novel laryngeal-imagery paradigm, we can establish the relative effectiveness of the novel paradigm. Machine learning algorithms were used to classify the EEG signals which had been transformed into frequency space (spectrograms) and common spatial pattern (CSP) dimensions. The SSVEP control task was able to be classified with better accuracy (62.5\%) than the no information rate of 50\% on the test set, but motor activity/imagery and laryngeal activity/imagery control tasks were not. Although the laryngeal methods did not produce accuracies above the no information rate, it is possible that with a larger amount of higher-quality data, this could prove otherwise. In the future, similar research should focus on reproducing the methods used here with better quality and more data.
\end{abstract}


\section{Introduction}
A brain-computer interface (BCI) is a computing system which is responsible for interpreting and extracting electrical signals from the brain to send to a functional output device \citep{Shih2012}. A BCI can also be a system which is used for sending information into the brain via electrical stimulation of some kind. BCIs currently offer practical benefits to individuals with severe epilepsy, ALS, stroke, locked-in syndrome, and paralysis \citep{Morrell1295, 6775293}. In one of its most common forms, BCI spellers, BCIs use actions such as looking at flashing lights or imagining arm movement to control letter selection systems  \citep{GUY20185, Sellers257re7}. This letter selection system is controlled by measurements of electrical activity of the individual's brain, often using implants or electroencephalography (EEG) sensors \citep{6775293, Rezeika2018}. The non-intuitive ways of making selections with these systems can be slow and confusing to control. However, BCI are capable of recognizing different kinds of subtle artifacts in the electrical activity of the brain which might be used to create more effective BCI spellers and systems. The reason BCI are capable of assisting individuals who lack motor control is that BCIs recognize activity in the brain as opposed to motor movements (e.g. pressing keys on a keyboard) in order to transfer information. If our goal is to increase the quality of life of motor-deficient individuals, then we should provide them with BCIs which are intuitive to use, such as a communication system based on activity in the speech centers of the brain as opposed to the motor or visual centers. BCI paradigms may be more easily integrated when the innate functionality of the neural source of control matches the functionality of the output device. Ideally, motor centers of the brain would drive motor BCI, and communication centers of the brain (e.g. the speech centers) would drive communication BCI. 

BCI spellers are the most common BCI application at the time of this writing, and there are three main types of electrical artifacts which BCI spellers look for: a P300 event-related potential (ERP), steady-state visual evoked potentials (SSVEP), and motor imagery (MI) \citep{Rezeika2018}. A P300 is an electrical artifact in response to a stimulus which increases in amplitude when greater attention is paid to the stimulus \citep{Shih2012}. A SSVEP is another kind of event-related potential which occurs in response to a consistent flashing visual stimulus \citep{Iscan2018}. It can be difficult to compare different BCI systems, because of the large numbers of differences between different BCIs. For this reason, it is best to compare BCIs on the basis of their accuracy at completing a task with discrete outcomes \citep{mowla2018bcihandbook}. The maximum accuracies of P300, a SSVEP, and a MI  BCI spellers as reported by \citet{Rezeika2018} were 99.7\%, 98.78\%, and 85\%, respectively. BCIs for motor imagery tasks, such as imagining using one's left or right hand, have yielded accuracies in the range of roughly 65\% to 95\% in the literature, but these systems often require long training periods for reliable use  \citep{quiles_low-cost_2020, irimia_high_2018}. The P300 and SSVEP control paradigms are also beneficial, because the P300 and SSVEP artifacts in EEG signals occur naturally and therefore these BCI do not require much training \citep{MCFARLAND2017194, Rezeika2018}. Although MI-based BCI tend to be less accurate and require more training, they are better for controlling more complex multidimensional output such as moving a cursor on a screen or a robotic arm \citep{MCFARLAND2017194}.

In order to improve the effectiveness of communication in BCIs this study proposes a new way of controlling BCI by using imagined humming (laryngeal motor imagination). The larynx is a muscular structure located in the throat which moves the vocal folds in the throat to form different vocal pitches \citep{10.3389/fnhum.2013.00237}. Using laryngeal motor imagination may improve BCI in the long-term by allowing individuals to utilize part of their brain already involved in voluntary speech: the laryngeal motor cortex (LMC) \citep{SIMONYAN201415}. This study aims to demonstrate comparable classification accuracy of this novel laryngeal MI paradigm in relation to the established classification methods: SSVEP and traditional MI. If the laryngeal MI-BCI performs significantly worse in terms of classification accuracy than the other methods in this study, then clearly laryngeal MI-BCI will have been shown to be less effective than established methods. If this laryngeal MI-BCI is comparably effective at providing a non-motor control mechanism, then this system could serve as a foundation for a more sophisticated speech prosthesis which would convert imagined speech into sound or text generated by electronics. This may be more intuitive for use than using other cognitive artifacts like converting imagined arm movement into speech, and could increase the quality of life for the individuals in need of speech prostheses as well as lead to the development of other related technologies.

\section{Methodology}
\label{sec:methodology}
\subsection{Materials and Tools}
This study used an OpenBCI EEG Mark IV headset with the Cyton, Daisy, and WiFi boards, which collects data from 16 EEG channels. Dry comb EEG electrodes with an Ag-AgCl coating were used with two ear clips as reference electrodes. An elastic strap that connects to the sides of the OpenBCI headset and sits under the participant's chin was used to ensure good sensor contact. Without this strap, good sensor contact could not be achieved. Data collection and processing was done with the OpenBCI GUI and Python with packages including brainflow, MNE, and pycaret. Data was collected using the brainflow Python package over a direct WiFi connection to a computer.

\begin{figure}
    \centering
    \includegraphics[scale=0.5]{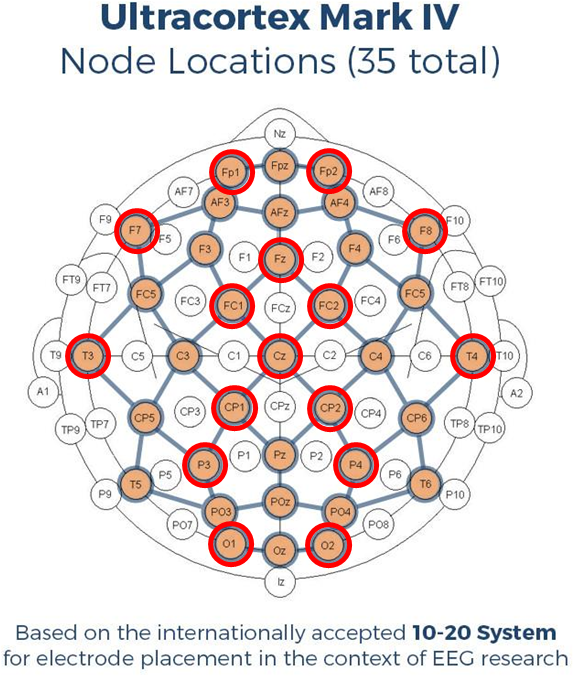}
    \caption{The 16 sensor locations that were used are shown as thick red circles. This is a top-down view where the anterior part of the brain faces the top of the page. The original image used as a template was from OpenBCI's website \citep{noauthor_ultracortex_nodate}.}
    \label{sensor-locations}
\end{figure}

The EEG sensor locations used the 10-20 system, with locations Fp1, Fp2, CP1, CP2, FC1, FC2, O1, O2, F7, F8, Fz, Cz, T3, T4, P3, P4 as shown in Figure~\ref{sensor-locations}. These were chosen to spread out the sensors while having sufficient sensors near the visual cortex for SSVEP/alpha wave detection and sensors near the motor cortex to detect motor activity and motor movement. Data was collected at a frequency of 1000Hz.

\subsection{Experiment Procedure}
\begin{figure}
    \centering
    \includegraphics[scale=1]{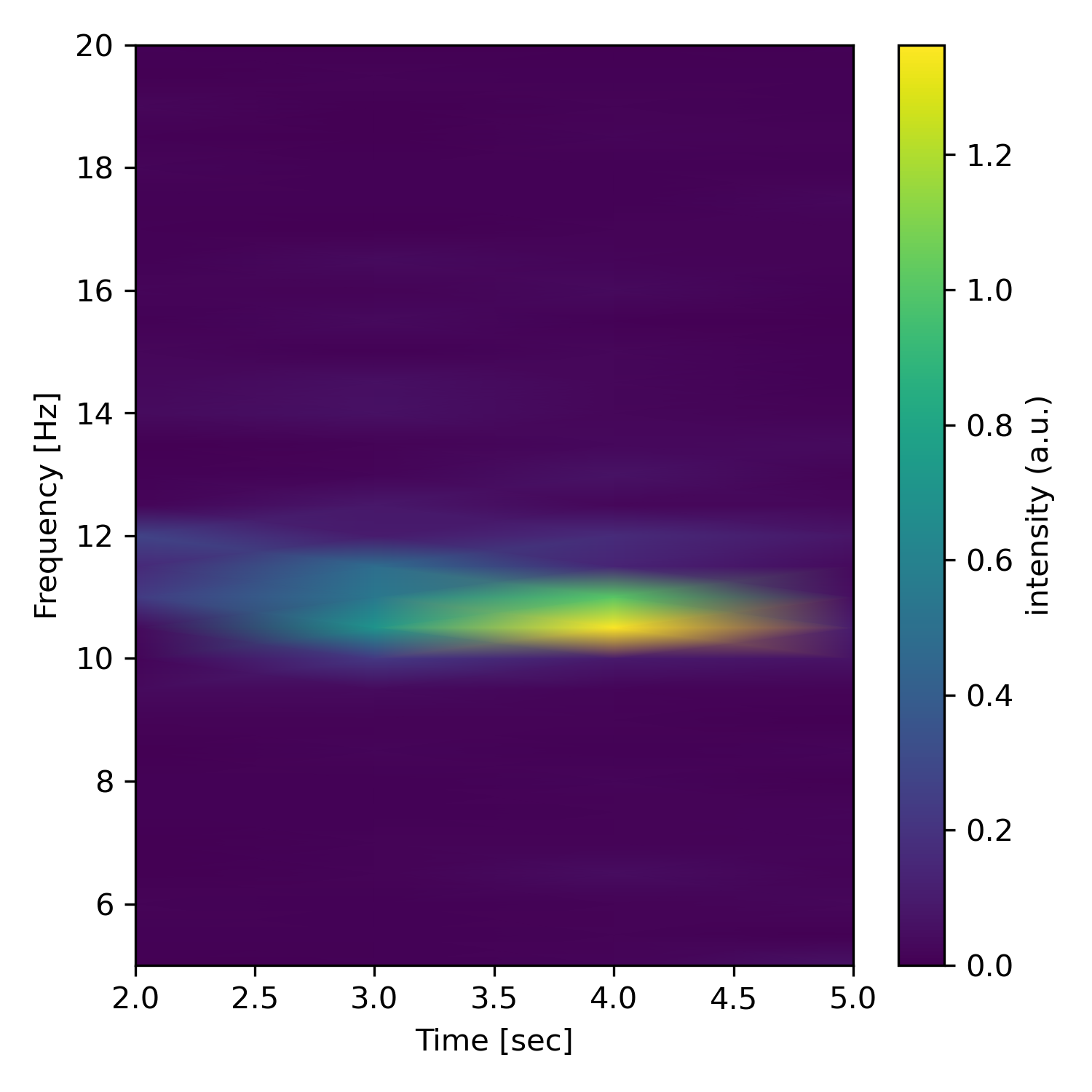}
    \caption{A spectrogram showing alpha waves that appear when a subject closes their eyes.}
    \label{alpha_waves}
\end{figure}

Participants were asked to read and sign an informed consent form approved by the Regis University IRB. The participant was instructed to wear an OpenBCI headset for recording EEG data and to face a computer screen where the experimental protocol was presented. The OpenBCI GUI was then used to ensure that sensors made sufficient contact with the scalp. This was done by examining the impedance of the EEG sensors, which should be at most 50 kOhm. While the participant closed their eyes for a few seconds, a live spectrogram using the OpenBCI GUI with the O1 and O2 EEG sensor channels was used to confirm that alpha waves (a peak around 10Hz; see Figure~\ref{alpha_waves}) appeared when the participant closed their eyes. Once the headset had been adjusted so that the sensor contact was sufficient, the experiment was initiated.

The experimental procedure was started by running the data collection software. After initial parameters for data collection were set (data storage location, EEG configuration) within the program, data collection was initiated. Instructions were shown to the participant explaining the procedures in the experiment. In the instructions, a warning was shown explaining that the steady state visually evoked potentials (SSVEP) portion of the experiment shows rapidly flashing squares on the screen and that people with epilepsy or other conditions that may result in harmful responses to flashing lights should end the experiment immediately by pressing the “x” key. Six sets of tasks, with 10 trials in each task, were then presented on the screen, in order: an eyes opened/closed task, an SSVEP task, a motor activity task, a motor imagery task, a laryngeal activity task (humming/rest), and a laryngeal imagery task (imagined humming/rest).

In the eyes open/closed task, instructions were shown to the participant explaining the task. A lower-pitched sound was then played, and the participant closed their eyes. After 5 seconds, a higher-pitched sound was played, and the participant opened their eyes for 5 seconds. This was repeated several times. An example of alpha waves from one of these trials is shown in Figure~\ref{alpha_waves}.

\begin{figure}
    \centering
    \includegraphics[scale=1]{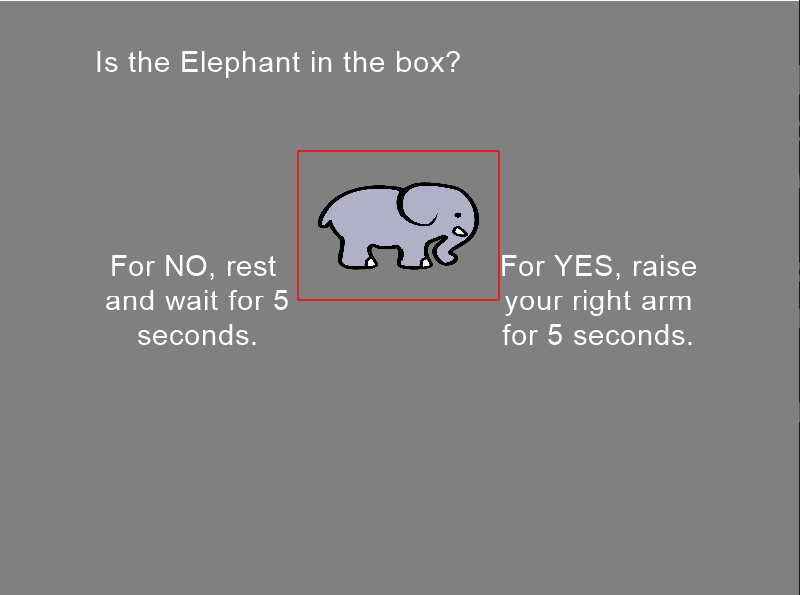}
    \caption{The screen capture shown above is an example from the experimental protocol GUI. The protocol gathers data pertaining to alpha waves, SSVEP, motor activity of the right arm, right arm motor imagery, laryngeal activity (humming) and laryngeal imagery (imagined humming).}
    \label{gui}
\end{figure}

In the next 5 sets of tasks, a cartoon picture of an elephant was shown to the user, either inside or outside of a box. If the elephant was inside the box, the participant should respond “yes” by pressing the right arrow key on the keyboard. If the elephant was outside the box, the user should respond “no” by pressing the left arrow key. If the participant pressed the wrong key, they were shown a warning and then the same configuration of the elephant and box was shown again until the participant pressed the correct key. The “yes” or “no” response then also corresponded to an SSVEP, motor action/imagery, or laryngeal action/imagery task. Ten trials were recorded in each of the 5 control paradigms. An example of this section of the experiment GUI is shown in Figure~\ref{gui}.

The second task used SSVEP control. Box-shaped visual stimuli (video files) were created to flash at frequencies of 10Hz and 15Hz. The SSVEP boxes were generated using sine waves so that the box would be completely white at the peak of the sine wave and completely black in the trough of the sine wave (with brightness proportional to the sine wave intensity in between peak and trough). These frequencies were chosen to have substantial separation to make signal detection easier, as well as being slow enough that conventional computer monitors could display them. The frequencies are also resonant with the 60Hz refresh rate common with computer monitors which helps ensure the correct frequencies will be shown. In the experiment, the user first pressed the correct key to signify the elephant was in or out of the box, then the two flashing squares were shown; one on the right of the screen (10Hz) and one on the left (15Hz). The flashing stimuli squares were presented approximately 4 in apart in 3.5 in x 3.5 in squares. The participant should look at the square on the right side of the screen to signify “yes” (the elephant was in the box) or look at the square on the left of the screen to signify “no”. After looking at the appropriate flashing square for 5 seconds, the next iteration of the SSVEP trial began. This was repeated 10 times.

The next task was actual motor activity. After the user pressed the correct key to signify the elephant was in or out of the box, the user was instructed to raise their right arm for “yes” or rest for “no” for a duration of 5 seconds. This was repeated 10 times. 

The next task was imagined arm movement. After the user pressed the correct key to signify the elephant was in or out of the box, the user was instructed to imagine raising their right arm for “yes” or rest for “no” for 5 seconds. This was repeated 10 times.

The next task was laryngeal motor actuation. After the user pressed the correct key to signify the elephant was in or out of the box, the user was instructed to make a humming sound for “yes” and rest for “no” for 5 seconds. This was repeated 10 times. 

The final task was laryngeal motor imagery. After the user pressed the correct key to signify the elephant was in or out of the box, the user was instructed to imagine making a humming sound for “yes” or rest for “no” for 5 seconds. This was repeated 10 times.

The data was then saved using the MNE Python package as .fif and Python pickle files. The data is available on the project's GitHub repository which can be found at  \href{Github}{https://github.com/nateGeorge/openbci\_laryngeal\_imagery} \citep{george_nategeorgeopenbci_laryngeal_imagery_2021}.

\subsection{Data preparation}
\begin{figure}
    \centering
    \includegraphics[scale=0.5]{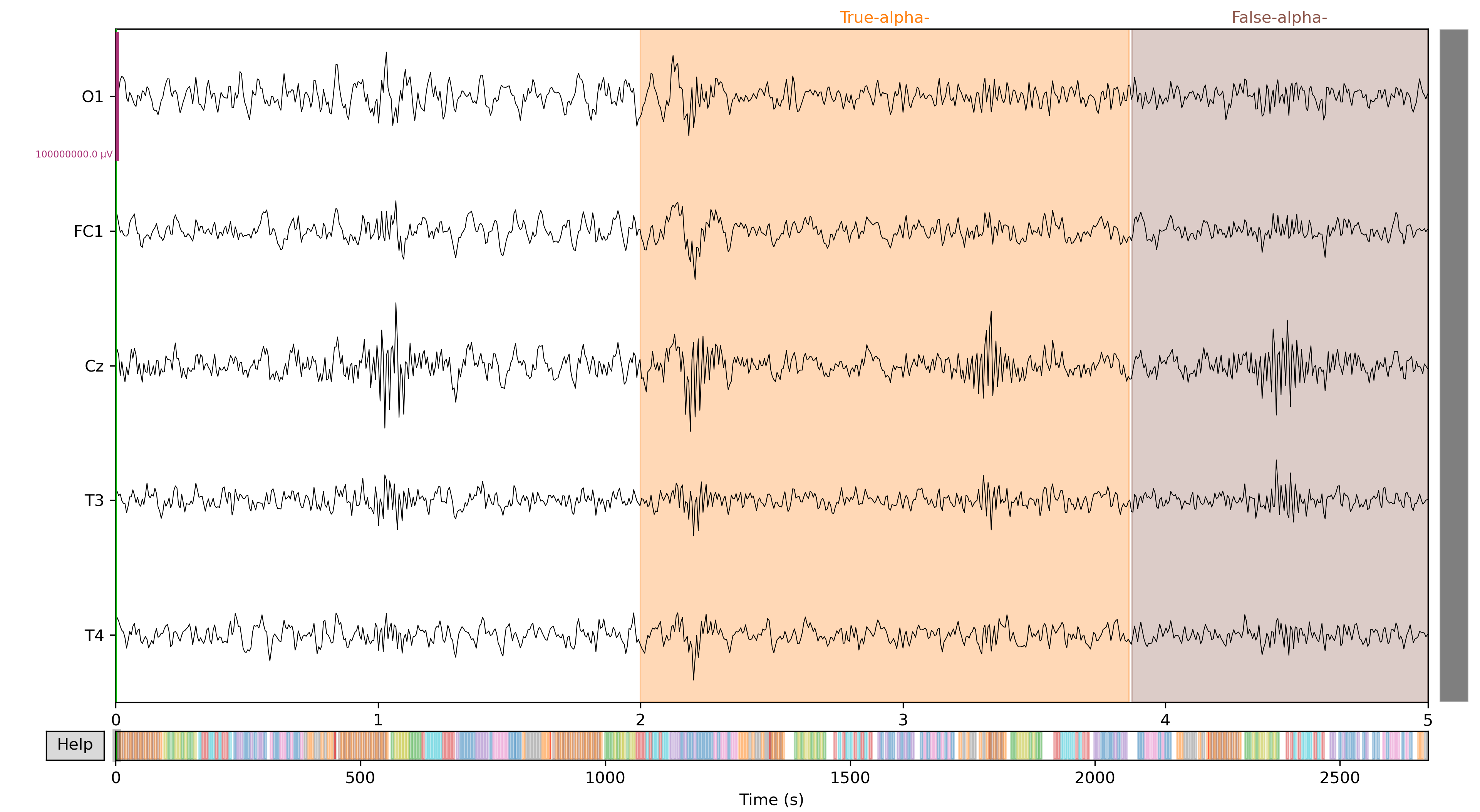}
    \caption{The above figure shows a 5 second window of EEG data from 5 of the 16 available sensors. The Cz channel shows some artifacts from the WiFi board on the OpenBCI headset that show up as periodic wavelets.}
    \label{data_sample}
\end{figure}

Before analysis, data was cleaned to remove noise. The first few seconds of OpenBCI data collected through brainflow and Python usually has a large spike in the signal, so the first 2 seconds of data from each experiment was removed. A bandpass filter from 5 to 50Hz was used, which removed signal drift, and a 60Hz notch filter was used to remove electrical noise from the surrounding electronics. Lastly,  standardization (subtraction of the mean and division by the standard deviation) was used separately on each channel to minimize the contribution from noisy channels (which had larger standard deviations). Another solution was tried where the noisy channels were flattened to 0, but standardization seemed to reduce noise in spectrograms more effectively. An example of the first few seconds of a few channels of the raw EEG data is shown in Figure~\ref{data_sample}. This figure demonstrates noise from the WiFi transceiver in this sample of data (periodic wavelets). Independent component analysis (ICA) was attempted as a way to remove this noise, but was unsuccessful.

To create SSVEP and alpha wave spectrograms, the channels O1, O2, P3, and P4 were used since these sensors were closest to the visual cortex where visual processing is most likely to be recorded \citep{MCFARLAND2017194}. For laryngeal activity and imagery spectrograms, all channels were used. For all spectrograms, 2000 datapoints (2s of data) were used in the window for each Fourier transform, and each window overlapped by 1000 datapoints (1s). A range of 5-50Hz was used for spectrogram features in machine learning algorithms. If the maximum frequency was not restricted to 50Hz and was instead 100Hz or 500Hz, the ML classifiers tended to use high frequencies as the most important features (near the top of the frequency range). This was confirmed by examining feature importances from a light gradient boosting model.

For spectrogram-based machine learning classification  (on the SSVEP and laryngeal tasks) spectrograms were created as described above. To prevent data leakage between train and validation/test sets, each 5s epoch of data was labeled as a group. Train and test datasets were created by taking a random set of 80\% of the data (by groups or 5 second continuous experimental sections) as the train set, and the remainder as the test set. This ensures the train and test sets have no overlapping data and no data possible leakage (i.e. no training data used in the validation or test sets). This data was then used with the pycaret auto machine learning Python package to try several machine learning models and select the best model by accuracy. The accuracies on the train set were calculated by 3-fold cross-validation (CV). Group labels in the train set were then re-labeled so that three distinct subsets remained in the training set. Accuracy of the best machine learning model was evaluated on the train and test sets. The best machine learning models with respect to each of the five tasks are shown in Table~\ref{accuracies}.

\begin{table}[]
\centering
\begin{tabular}{|l|l|l|l|}
\hline
Task                                              & Model                                      & \thead{Train set\\ accuracy (\%)} & \thead{Test set\\ accuracy (\%)} \\
\hline
\makecell{SSVEP\\ (look at 10Hz or 15Hz square)}               & \makecell{Logistic regression\\with L2 regularization} & 72.4               & 62.5              \\
\hline
\makecell{Motor activity \\ (raise right arm or rest)}          & Decision tree                                   & 81.2               & 40.0              \\
\hline
\makecell{Motor imagery\\ (imagine raising right arm or rest)} & Linear SVM                        & 41.2               & 60.7              \\
\hline
\makecell{Laryngeal activity \\ (hum or rest) \\ CSP features}                  & Linear SVM                        & 65.4               & 42.2              \\
\hline
\makecell{Laryngeal imagery\\ (imagine humming or rest) \\ CSP features}       & Linear SVM                          & 56.0               & 47.6   \\
\hline
\makecell{Laryngeal activity \\ (hum or rest) \\ Spectrogram features}                  & \makecell{Quadratic discriminant\\ analysis}                       & 57.3               & 48.0              \\
\hline
\makecell{Laryngeal imagery\\ (imagine humming or rest) \\ Spectrogram features}       & k-nearest neighbors                          & 65.4               & 50.0   \\
\hline
\end{tabular}
\caption{Binary tasks with the best-performing machine learning model and accuracy scores on the train and test sets (80\% of data was in the training set).}
\label{accuracies}
\end{table}

For motor activity/imagery tasks and laryngeal activity/imagery tasks, common spatial pattern (CSP) transforms were used. The CSP fit and transform was performed on the train set, and the test set was transformed by the CSP operation. Train and test sets were constructed in the same way as the SSVEP data, using group labels for continuous 5s data chunks and using separate groups in the train and validation/test sets. Again, pycaret was used to select the best performing machine learning model by accuracy. Accuracy was evaluated on the train and test sets.

\subsection{Participants and Datasets}
Two participants (the paper's authors) were used for data collection: both males aged 21 and 34 at the time of the experiment. Both participants had substantial amounts of hair which made dry sensor contact more difficult. Three datasets from each participant were collected for a total of 6 experiments. Several alpha wave epochs were collected for each experiment (at least 5). Ten epochs for each of the five control paradigms (SSVEP, motor activity, motor imagery, laryngeal activity, laryngeal imagery) were collected, except for the last experiment, which had 10 of each of the control paradigms but only 3 of the laryngeal imagery epochs due to a misconfiguration. This means that for analysis there were 60 trials in each of the control paradigms (with 30 of each of the "yes" and "no" sections) except for the laryngeal imagery which had 53 trials.

\section{Results and Discussion}
\begin{figure}
    \centering
    \includegraphics[scale=1.2]{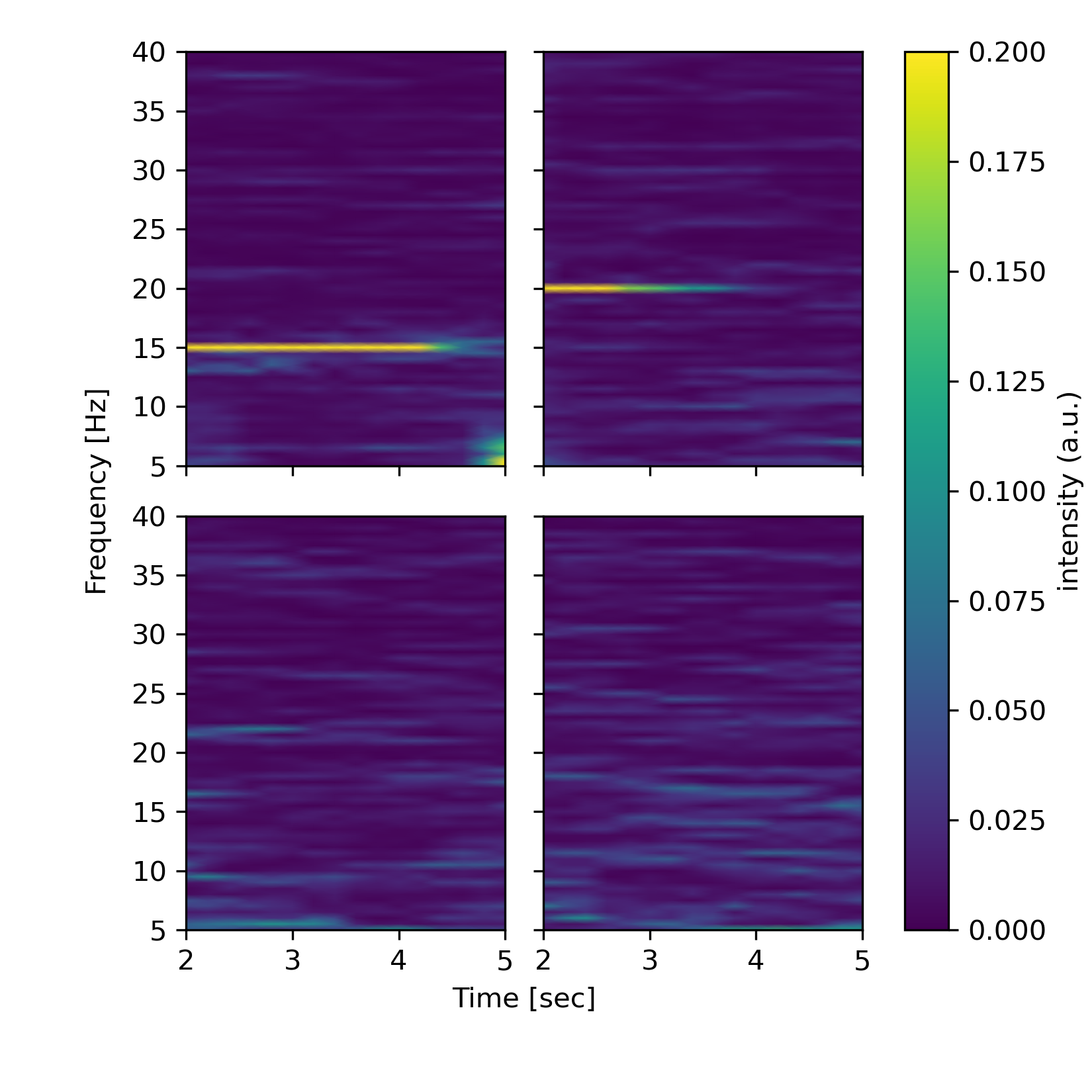}
    \caption{The left column of spectrograms show 15 Hz SSVEP signals, while the right column of spectrograms show 10 Hz SSVEP signals. The top row of spectrograms shows data with SSVEP signals present, and the bottom row of spectrograms shows examples when the SSVEP signals were not present.}
    \label{ssvep_spectrograms}
\end{figure}

The first experiment in the series was SSVEP, where a flashing square with a frequency of 15Hz was shown on the left (for a selection of "no") and a flashing square with a frequency of 10Hz was shown on the right (for a selection of "yes"). After pre-processing, spectrograms were created to visualize quality of the data. Ideally, sharp signals at the frequency of the flashing stimulus (and their harmonic frequencies such as 20Hz and 30Hz) would be present, and the signal at the frequency that the participant looked at would be strongest \citep{Rezeika2018, 6775293}. As can be seen in Figure~\ref{ssvep_spectrograms}, some data had clear SSVEP signals present while others did not. In fact, the majority of experimental sections did not have clear SSVEP signals visible from spectrograms. Additionally, a few spectrograms that should have had the 15Hz signal showed the 10Hz signal, and vice versa. It is not clear why this happened. We can also see from Figure~\ref{ssvep_spectrograms} that the second harmonic (20Hz) signal was clearly visible where the 10Hz signal was shown, and this second harmonic was stronger than the 10Hz signal itself. This was true in most of the 10Hz SSVEP signals, but not the 15Hz signals.

After using pycaret to select the best machine learning algorithm with the labeled spectrogram data, a logistic regression model was found to perform best with 72.4\% accuracy on the train set and 62.5\% accuracy on the test set. The accuracies beat the no information rate of 50\%, but were not high enough to provide for a reliable information output channel. This is likely due to noisy data and the issue of 15 and 10Hz signals in the spectrogram analysis occasionally showing up in the wrong epochs (for example, a 10Hz signal showing up where 15Hz is expected).

The motor/laryngeal activity and imagery data was processed with CSP (common spatial pattern) transforms and machine learning performed on these CSP transforms. The laryngeal activity and imagery data was also processed in the same way as the SSVEP data by creating spectrograms from all channels. The reason for trying a spectral approach with the laryngeal but not the motor action/imagery data was that it was not clear if the features from making and imagining sounds would show up in the spatial or spectral domain of the EEG data. The best model for each control paradigm was selected with pycaret autoML, and the results of the best ML models for each scenario are shown in Table~\ref{accuracies}.

We can see that ML categorization of these motor and laryngeal tasks had similar accuracies on the train and test sets, with none of them beating the no information rate of 50\% except for motor imagery.  This indicates EEG signals from both the motor and laryngeal control paradigms could not be successfully classified in this study. Although the motor imagery shows an accuracy comparable to SSVEP and above the 50\% random guessing baseline, the accuracy on the train set was only 41.2\%, decreasing confidence that the result on the test set is meaningful. Additionally, we would expect the accuracy for motor activity to be greater than that of motor imagery, but it had an accuracy 20.7\% less than the motor imagery paradigm. The results here are near the lower end of the range of 65\% to 95\% accuracy for classification of left versus right hand movements via dry EEG electrodes in the literature \citep{quiles_low-cost_2020, irimia_high_2018}.

Many of the best models for motor and laryngeal control paradigms ended up being linear support vector machines (SVMs), although decision tree, quadratic discriminant analysis, and k-nearest neighbors models were also found to perform best for some of the specific control paradigms and data preparation types (FFT spectrograms and CSP filters).

The low ML classification accuracies of motor and laryngeal control are likely due to noisy data and using a small amount of data. Data quality may be low due to the use of dry sensors, using the OpenBCI WiFi transmitter which seemed to cause artifacts in the data, and electrical noise (e.g. from computers, cell phones, and other electronics in the room at the time of data collection). It is expected that using the Bluetooth transmitter and saving data to an on-board SD card could improve data quality, or other electrical modifications to the OpenBCI circuit board. More data could be collected by repeating the experiment or using a different headset which allows for more sensors to be used. Using wet sensors should improve data quality as well.

\begin{figure}
    \centering
    \includegraphics[scale=0.3]{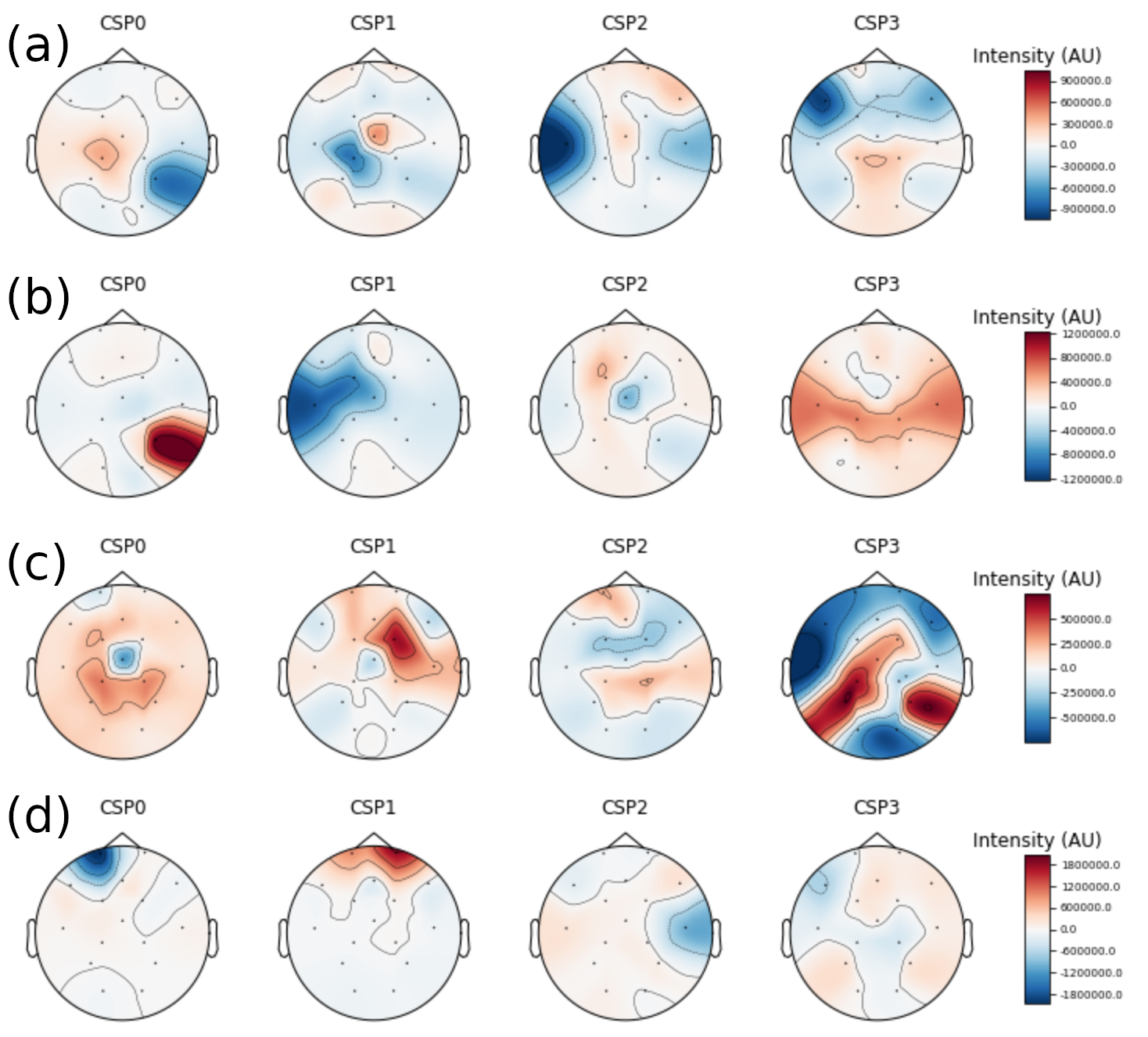}
    \caption{This figure shows the common spatial pattern (CSP) transforms of (a) motor activity (raising one's right arm), (b) motor imagery (imagining raising one's right arm), (c) laryngeal activity (humming), and (d) laryngeal imagery (imagining humming). This is a top-down view of the activity of the head with the anterior of the brain facing the top of the page.}
    \label{csp_filters}
\end{figure}

Since the imagery and movement tasks used CSP filters to preprocess data, we can examine the results of the CSP transforms to look for trends in location of motor and laryngeal activity. Although the data seemed noisy in general, the amount of noise was not constant, even in a single experiment. This was observed by looking at spectrograms of different sections of the experiment. So we may still be able to glean information from the CSP-transformed data. Figure~\ref{csp_filters} shows the filters for motor activity, motor imagery, laryngeal activity, and laryngeal imagery. The CSP filters are ordered from left to right by order of magnitude and importance.  The first two CSP filters demonstrate that motor activity strongly centers around the C3, CP1 and FC2 10-20 EEG sensor locations. This is in line with other findings and the location of the motor cortex \citep{chen_mi}. We also see a strong signal around the T6 sensor location (on the back right of the head), which is bordered by sensors O2, P4, and T4. This may be in part due to noise on some of the experiments in the P4 sensor.

For the motor imagery, we see similar but weaker results. The same T6 spot shows activity, but the second CSP filter has a strong signal on the left side of the head from the C3 sensor down to the T3 sensor. The CSP highlighting the left side of the brain make sense for our particular motor action and imagery (imagining or actually raising the right arm), because control of lateralized movements of the body occur in the contralateral hemisphere of the brain. However, the CSP signal around the T6 sensor location could be due to noise in the data or the small dataset size. The CSP results give hope that data quality was enough to see some evidence of signal where it was expected, even though this was not enough to accurately classify the data using ML.

For laryngeal action and imagery, very different areas were highlighted by the CSP filters. In the case of laryngeal action, we  see activity near the Cz sensor near the motor strip as we would expect (the pre-motor areas near the laryngeal motor cortex, or LMC). The FC2 sensor also shows a signal in the second CSP filter. However, the CSP transforms show very little activity in the motor and pre-motor areas in the case of laryngeal imagery. Laryngeal imagery seemed to produce greater activity in the pre-frontal cortex near the Fp1 and Fp2 sensors. This is consistent with the pre-frontal cortex's role in executive decision making and working memory, since these are often associated with the prefrontal cortex \citep{Koechlin7651}. This may be evidence for planning movement of the larynx, but this does not necessarily allow for classifying laryngeal imagination as distinct from any other form of planning.

Since the performance of the ML algorithms on the motor and laryngeal data was no better than random chance, we should not give too much weight to the results of these CSP transforms, although they could be used as a basis for future studies.

\section{Conclusions}
EEG data was collected to evaluate a novel BCI control paradigm, imagined laryngeal activity.  This was compared to other BCI control paradigms, and all paradigms were used to signal a binary choice of yes or no. The performance of one optimized machine learning classifier for each of the paradigms is presented in Table~\ref{accuracies}. This includes paradigms used for reference such as motor activity and laryngeal activity. Due to the unknown nature of the laryngeal EEG signature, laryngeal activity and imagination were analyzed in the spatial domain (CSP) and the frequency domain (FFT spectrograms).
The established paradigms, SSVEP and MI were found to have accuracies of 62.5\% and 60.7\% when analyzed on the test set using a logistic regression with L2 regularization classifier and a linear support vector machine (SVM), respectively.

Laryngeal activity and imagery classification was done on the spatial domain (CSP), with the best models resulting in linear SVMs and on the frequency domain using Quadratic discriminant analysis and k-nearest neighbors classifiers. These classification methods yielded accuracies of 42.2\% (spatial-domain laryngeal activity), 47.6\% (spatial-domain laryngeal imagery), 48.0\% (frequency-domain laryngeal activity), and 50.0\% (frequency-domain laryngeal imagination). None of these paradigms, even the control paradigms where laryngeal activity was present, were able to overcome the 50\% no information rate. This seems to add evidence that significant noise was present in the data. This can be improved upon by using wet electrodes, taking measures to remove noise from the WiFi OpenBCI circuit board, and by simply collecting a larger data set.

CSP filters showed that motor and laryngeal activity tended to result in EEG signals over the motor cortex consistent with functional neuroanatomy. However, in the case of laryngeal imagery, activity seemed more intense in the pre-frontal cortex. Due to the failure to establish high accuracies in the established BCI paradigms, this experiment can draw no conclusion about the efficacy of the imagined laryngeal signal for its use as a BCI control paradigm. It is still unknown if imagined laryngeal activity can be detected from EEG signals. Future research with more and higher quality data is needed to answer the question of whether laryngeal imagination can be used as a control mechanism with EEG signals.

\bibliographystyle{unsrtnat}
\bibliography{references}  






\end{document}